\documentclass[11pt,a4paper,colorlinks]{article}
\usepackage{jheppub}
\relax

\def\bes{\begin{eqnarray}}
 \def\ees{\end{eqnarray}}
\def\be{\begin{equation}}
\def\ee{\end{equation}}
\def\bs{\begin{subequations}}
\def\es{\end{subequations}}
\newcommand{\een}{\end{subequations}}
\newcommand{\ben}{\begin{subequations}}
\newcommand{\beq}{\begin{eqalignno}}
\newcommand{\eeq}{\end{eqalignno}}

 \def\lx{\lambda}

\def\Dt{{\tilde{\Delta}}}

 \def\Dr{\Delta r}
 \def\Dt{\Delta t}
 \def\Ac{{\mathcal A}}
 \def\Bc{{\mathcal B}}
 \def\Cc{{\mathcal C}}
\def\Dc{{\mathcal D}}

\def \lta {\mathrel{\vcenter
     {\hbox{$<$}\nointerlineskip\hbox{$\sim$}}}}

\title{Dynamical classicalization}
\author[1]{J. Rizos}
\author[2]{and N. Tetradis}
\affiliation[1]{Department of Physics,\\ University of Ioannina, Ioannina 45110, Greece}
\affiliation[2]{Department of Physics,\\ University of Athens, Zographou 15784, Greece}

\emailAdd{irizos@uoi.gr}
\emailAdd{ntetrad@phys.uoa.gr}

\abstract{
We integrate numerically the nonlinear equation of motion for a collapsing spherical wavepacket in the context of theories that
are expected to display behavior characteristic of classicalization.
The classicalization radius sets the scale for the onset of significant deformations of the
collapsing configuration, which result in the formation of shock fronts.
A characteristic observable feature of the classicalization process is the creation of an outgoing
field configuration that extends far beyond the classicalization radius. This feature
develops before the deformed wavepacket reaches distances of the order of the
fundamental scale.
We find that in some models the scattering problem may not have
real solutions over the whole space at late times. We determine the origin of this behavior and
discuss the consistency of the underlying models.
}


\arxivnumber{arXiv:1112.5546 [hep-th] }
\begin{document}
\maketitle

\section{Introduction}\label{intro}

The physics  of classicalization concerns the nature of high-energy scattering in certain classes of nonrenormalizable field theories.
The proposal of refs. \cite{dvalex1,dvpirts,dvali,dvalex2} is that scattering can take place at length scales much larger than the
typical scale associated with the nonrenormalizable terms in the Lagrangian. The behavior is similar to ultra-Planckian
scattering in gravitational theories, during which a black hole is expected to start forming at distances comparable to the
Schwarzschild radius. For sufficiently large center-of-mass energy, the Schwarzschild radius can be much larger than the
Planck length. If the formation and subsequent evaporation of a black hole are viewed as parts of a scattering process, it becomes
clear than the physical length scale determining the cross section can be the Schwarzschild radius and not the Planck scale.
It has been advocated that similar behavior may be observed
in some nonrenormalizable field theories \cite{dvalex1,dvpirts,dvali,dvalex2}.
In scattering processes, the center-of-mass energy introduces a new dimensionful scale in the problem that can be
used to define the analogue of the Schwarzschild radius. This new length scale is termed classicalization radius. It
defines the distance at which scattering starts and, for sufficiently high center-of-mass energy,
can be much larger than the fundamental length scale $L_*$ of the theory, determined by the coefficient of the non-renormalizable
term in the action. The scenario has very interesting collider phenomenology \cite{lhc}.

An idealized scattering process was proposed in refs. \cite{dvpirts,dvali} as a means to check the validity of the
classicalization arguments. The process involves an initial spherical wavepacket of very large radius that propagates towards
the center of symmetry. The solution of the classical equation of motion can provide a clear indication of the onset of scattering.
One expects that it should be possible to verify that this starts
at length scales much larger than $L_*$. An approximate analytical
solution of the equation of motion was presented in refs. \cite{dvpirts,dvali}. It demonstrates that there are significant
corrections to the initial wavepacket at length scales of order $r_*$, with the simultaneous creation of a long tail in the
field configuration, stretching out to large distances. However, an exact solution of the equation of motion is still lacking.
In ref. \cite{brt}, it was suggested that the equation of motion can be written as a quasi-linear
partial differential equation of second order. For the initial spherical wavepacket this equation is hyperbolic,
supporting the propagation of wavelike configurations. However, when the center of the wavepacket approaches $r_*$,
the equation may become elliptic in certain regions. This change could result in the suppression of wave propagation
and the emergence of scattering.

In this work we present numerical solutions of the field equations of motion for two theories, described by nonrenormalizable
Lagrangians, which are expected to generate behavior characteristic of classicalization. We discuss in detail how
the typical features of classicalization emerge during the evolution of the idealized spherical wavepacket described above.
We also point out several novel features that seem to be inherent in scattering processes in the context of such theories.
In the following two sections we describe the theories in question.  In section 4 we present the method we employ for the
integration of the equations of motion. We also discuss the crucial issue of existence or absence of real solutions
in certain ranges of the radial coordinate $r$. In section 5 we present the results of the numerical integration.
In the final section we discuss the interpretation of our results.

\section{Quartic action}\label{quartic}

The prototype scalar theory in which classicalization is expected to occur is described by the Lagrangian density
\be
{\cal L}=\frac{1}{2} \left(\partial_\mu\phi \right)^2 -\delta_1 \frac{L^4_*}{4}\left( \left(\partial_\mu\phi \right)^2\right)^2.
\label{lagrangian0} \ee
We allow for both signs of the higher derivative term by assuming that $\delta_1$  can take the values
$\delta_1=\pm 1$.
The equation of motion of the field $\phi$ is
\be
\partial^\mu\left[  \partial_\mu \phi \left(1-\delta_1 L_*^4\left(\partial_\nu\phi \right)^2\right)\right] =0.
\label{eomi} \ee

An idealized scattering process, which displays the characteristic behavior associated with classicalization, involves
a collapsing spherical wavepacket initially represented by a Gaussian
of width $a$ centered around a radius $r_0$. It has the form
\be
\phi_0(t,r)=\frac{A}{r}\exp\left[-\frac{\left(r+t-r_0\right)^2}{a^2} \right].
\label{wave} \ee
Using perturbation theory, it was shown in refs. \cite{dvpirts,dvali} that
the field configuration is strongly deformed when the peak of the wavepacket reaches the
classicalization radius
\be
r_*\sim L_* \left( \frac{A^2 L_*}{a}\right)^{1/3}.
\label{clr} \ee
This can be substantially larger than the fundamental length scale $L_*$ when the center-of-mass energy
$s\sim A^2/a$ is much larger than $1/L_*$.
It must be noted that the analysis of refs. \cite{dvpirts,dvali} was carried out for
the case $\delta_1=-1$. However, it is straightforward to check that the same arguments apply to the case
$\delta_1=1$ as well.

Another point of view was presented in ref. \cite{brt}. It was argued that the behavior associated with classicalization is related to the
dynamical change of type of the classical equation of motion. For a spherically symmetric
configuration, eq. (\ref{eomi}) takes the form
\be
\left(1-3 \lx \phi_t^2+\lx\phi_r^2 \right)\phi_{tt}-\left(1- \lx \phi_t^2+3\lx\phi_r^2 \right)\phi_{rr}
+4\lx \phi_r\phi_t \,\phi_{tr}=\frac{2\phi_r}{r}\left(1- \lx \phi_t^2+\lx\phi_r^2 \right),
\label{eomsph0} \ee
where $\lx=\delta_1 L^4_*$ and subscripts denote partial derivatives.
 The above equation can be
expressed as a conservation law, in the form
\be
\partial_t\left[\phi_t \left(1- \lx \phi_t^2+\lx\phi_r^2 \right)\right]
-\frac{1}{r^2}\partial_r\left[r^2\phi_r \left(1- \lx \phi_t^2+\lx\phi_r^2 \right)\right]
=0.
\label{eomcons} \ee
It can also be expressed as
\be
\partial_t\left[\frac{1}{2} \left(\phi_t^2+\phi_r^2 \right)+
\frac{\lx}{4} \left(-3 \phi_t^4+ \phi_r^4+2 \phi_t^2\phi_r^2  \right)\right]
-\frac{1}{r^2}\partial_r\left[r^2\phi_t \phi_r \left(1- \lx \phi_t^2+\lx\phi_r^2 \right)\right]
=0.
\label{eomcons2} \ee
This last equation is equivalent to the conservation of the energy-momentum tensor resulting from the
Lagrangian of eq. (\ref{lagrangian0}).
 The quantity
\be
\rho=\frac{1}{2} \left(\phi_t^2+\phi_r^2 \right)+
\frac{\lx}{4} \left(-3 \phi_t^4+ \phi_r^4+2 \phi_t^2\phi_r^2  \right)
\label{density0} \ee
is the local energy density.
The initial conditions for the solution of the equation of motion are of the Cauchy type:
\be
\phi(0,r)=\phi_0(0,r),
~~~~~~~~~~~~~~~~~~~~~~~
\partial_t\phi(0,r)=\partial_t \phi_0(0,r),
\label{init} \ee
where we assume that $r_0$ is much larger than any other physical scale.

The properties of eq. (\ref{eomsph0}) become more transparent if it is written in the form of a quasilinear second-order partial
differential equation:
\be
\Ac(\phi_t,\phi_r) \, \phi_{tt}+ \Bc(\phi_t,\phi_r) \, \phi_{tr}+ \Cc(\phi_t,\phi_r) \, \phi_{rr}= \Dc(\phi_t,\phi_r,r),
\label{pdeform} \ee
with
\begin{eqnarray}
\Ac(\phi_t,\phi_r)&=&1-3 \lx \phi_t^2+\lx\phi_r^2
\label{aai} \\
\Bc(\phi_t,\phi_r)&=&4\lx\phi_t\phi_r
\label{bbi} \\
\Cc(\phi_t,\phi_r)&=&-\left(1- \lx \phi_t^2+3\lx\phi_r^2\right)
\label{cci} \\
\Dc(\phi_t,\phi_r,r)&=&\frac{2\phi_r}{r}\left(1- \lx \phi_t^2+\lx\phi_r^2 \right).
\label{ddi} \end{eqnarray}
The type of this partial differential equation is determined by the discriminant
\be
\Delta=\frac{1}{4}(\Bc^2-4\Ac\Cc)=3\left(\frac{1}{3}-\lx \phi^2_t+\lx \phi_r^2 \right)\left(1-\lx \phi^2_t+\lx \phi_r^2 \right).
\label{discr0} \ee
For $\Delta>0$ the equation is hyperbolic,
for $\Delta=0$ parabolic, while for $\Delta<0$ elliptic.
Hyperbolic equations admit wave-like solutions, while elliptic ones do not
support propagating solutions. The observation of ref. \cite{brt}
is that classicalization may be associated with the change of the type of the equation during the evolution of
the initial configuration.
It can be shown \cite{brt} that, if $\Ac$, $\Bc$, $\Cc$ are evaluated for the configuration (\ref{wave}),
the discriminant (\ref{discr0}) switches sign in a certain $r$-interval when the wavepacket moves into the vicinity of the classicalization
radius (\ref{clr}). The sign change occurs for both values of $\delta_1$. As a result, we expect
behavior associated with classicalization in both cases.

In the following sections we focus on the details of the evolution of an initial field configuration given by eq. (\ref{wave}).
We have managed to obtain only a numerical solution of
the equation of motion with this initial condition, which we present in section 5.
However, analytical solutions of the quasilinear second-order partial
differential equation (\ref{eomsph0}) can also be found and they are of interest, even though they do not satisfy
the initial condition (\ref{wave}).

 For $\phi=\phi(r)$,  eq. (\ref{eomcons}) gives
\be
\phi_r(1+\lx \phi^2_r)=\frac{c}{r^2},
\label{stat1} \ee
with $c$ a constant of integration.
If we are interested in a localized configuration of finite energy, we must select the root of the above equation for which
$\phi_r$  approaches $c/r^2$ for large $r$.
The solution with this asymptotic behavior extends down to $r=0$ for $\lx>0$.
 For $\lx<0$ the solution displays
a square-root singularity at a certain value $r_s\not= 0$. It is possible to join smoothly this solution with a second root of
eq. (\ref{stat1})  that again
extends from $r_s$ to infinite $r$. In this way the field $\phi$ becomes a double-valued function of position.
A similar construction was employed in ref. \cite{gibbons} in order to describe Dirichlet $(d-1)$-branes embedded in
$(d+1)$-dimensional Minkowski space, with the field $\phi$ corresponding to the transverse coordinate.
However, it is difficult to find an interepretation of such a solution within the framework of the four-dimensional scalar field
theory that we are considering: The field $\phi$ would be double-valued for $r>r_s$, while it would not be defined for
$r<r_s$.

The partial differential equation (\ref{eomsph0}) also has exact dynamical solutions of the form $\phi=\phi(z)$, with
$z=r^2-(t-t_0)^2$. They are given by
\begin{eqnarray}
\phi_t(t,r)&=&-2(t-t_0)\, h(r^2-(t-t_0)^2)
\label{ex1} \\
\phi_r(t,r)&=&2r\, h(r^2-(t-t_0)^2),
\label{ex2}
\end{eqnarray}
with $h(z)$  satisfying the differential equation
\be
z(1+12 \lx\, z\, h(z)^2) h'(z)+12 \lx\, z\, h(z)^3+2h(z)=0.
\label{hz} \ee
This equation has three branches of solutions, similarly to
eq. (\ref{stat1}). They correspond to the roots of the cubic equation
\be
4\lx z^2h^3(z)+z^2h(z)+c=0,
\label{cub} \ee
with $c$ an integration constant.
We do not present the explicit expressions because of their complexity. In the following
section, we discuss
in detail similar solutions that can be obtained in the context of a theory of the Dirac-Born-Infeld type.

\section{Dirac-Born-Infeld action}\label{eomm}

A related theory, which can also display the classicalization phenomenon, is described by a Lagrangian density of the
Dirac-Born-Infeld (DBI) type:
\be
{\cal L}=-\frac{1}{\delta_2 L_*^4}\sqrt{1-\delta_2 L_*^4\left(\partial_\mu\phi \right)^2},
\label{lagrangian} \ee
with $\delta_2=\pm 1$, as in the previous case.
The equation of motion of the field $\phi$ is
\be
\partial^\mu\left[ \frac{  \partial_\mu \phi}{\sqrt{1-\delta_2 L_*^4\left(\partial_\nu\phi \right)^2}}\right] =0.
\label{eom} \ee
The leading terms in the expansion of the l.h.s. of the above equations reproduce the quartic model of the previous
section, provided that we take $\delta_2=-2\delta_1$. The Lagrangian (\ref{lagrangian}) also includes a
cosmological constant term, which is irrelevant for our considerations.
In general, we expect similar behavior for the solutions of eqs. (\ref{eomi}), (\ref{eom})
if $\delta_1$, $\delta_2$ are taken with opposite signs. However, the higher-order terms resulting from the
expansion of the l.h.s. of eq. (\ref{eom}) are also important for the dynamics.

When expressed in spherical coordinates, eq. (\ref{eom}) can be put in the form
\be
\left(1+\lx\phi_r^2 \right)\phi_{tt}-\left(1- \lx \phi_t^2 \right)\phi_{rr}
-2\lx \phi_r\phi_t \,\phi_{tr}=\frac{2\phi_r}{r}\left(1- \lx \phi_t^2+\lx\phi_r^2 \right),
\label{eomsph} \ee
where $\lx=\delta_2 L^4_*$ and subscripts denote partial derivatives.
It is important to emphasize that, in order to obtain the above equation, eq. (\ref{eom}) must be multiplied by
$\left(1- \lx \phi_t^2+\lx\phi_r^2 \right)^{3/2}$. As a result, we must assume that
\be
1- \lx \phi_t^2+\lx\phi_r^2 \geq 0.
\label{constraint} \ee
This assumption is an obvious constraint imposed by the form of the Lagrangian density (\ref{lagrangian}).
Eq. (\ref{eomsph}) can also be
expressed as a conservation law, in the form
\be
\partial_t\left[ \frac{\phi_t }{\sqrt{1- \lx \phi_t^2+\lx\phi_r^2 }} \right]
-\frac{1}{r^2}\partial_r\left[r^2\frac{\phi_r }{\sqrt{1- \lx \phi_t^2+\lx\phi_r^2 }}\right]
=0.
\label{eomconsd} \ee
It can also be written in the form
\be
\partial_t\left[ \frac{1+\lx\phi_r^2 }{\lx\sqrt{1- \lx \phi_t^2+\lx\phi_r^2 }} \right]
-\frac{1}{r^2}\partial_r\left[r^2\frac{\phi_r\phi_t }{\sqrt{1- \lx \phi_t^2+\lx\phi_r^2 }}\right]
=0.
\label{eomconsen} \ee
This last equation is equivalent to the conservation of the energy-momentum tensor resulting from the
Lagrangian density of eq. (\ref{lagrangian}). The quantity
\be
\rho= \frac{1+\lx\phi_r^2 }{\lx\sqrt{1- \lx \phi_t^2+\lx\phi_r^2 }}-\frac{1}{\lx}
\label{density} \ee
is the local energy density, after the subtraction of a cosmological constant included in the Lagrangian.

Eq. (\ref{eomsph}) can be written in the form (\ref{pdeform})
with
\begin{eqnarray}
\Ac(\phi_t,\phi_r)&=&1+\lx\phi_r^2
\label{aa} \\
\Bc(\phi_t,\phi_r)&=&-2\lx\phi_t\phi_r
\label{bb} \\
\Cc(\phi_t,\phi_r)&=&-\left(1- \lx \phi_t^2\right)
\label{cc} \\
\Dc(\phi_t,\phi_r,r)&=&\frac{2\phi_r}{r}\left(1- \lx \phi_t^2+\lx\phi_r^2 \right).
\label{dd} \end{eqnarray}
The type of this partial differential equation is determined by the discriminant
\be
\Delta=\frac{1}{4}(\Bc^2-4\Ac\Cc)=1-\lx \phi^2_t+\lx \phi_r^2.
\label{discr} \ee
For $\Delta>0$ the equation is hyperbolic,
for $\Delta=0$ parabolic, while for $\Delta<0$ elliptic.
The DBI theory represents a special case. Requiring the Lagrangian density to be real
results in the constraint (\ref{constraint}), which excludes the possibility $\Delta <0$.
We can again estimate the distance $r$ at which eq. (\ref{pdeform}) changes type if $\Ac$, $\Bc$, $\Cc$ are
evaluated for the configuration (\ref{wave}). For both values of $\delta_2$, the discriminant (\ref{discr}) may vanish when the
wavepacket approaches a region in which $r$ is approximately given by eq. (\ref{clr}) \cite{brt}.

Exact analytical solutions of the evolution equation can be derived, similarly to the previous section.
 For $\phi=\phi(r)$, eq. (\ref{eomconsd}) gives
$\phi_r/\sqrt{1+\lx\phi^2_r}=\pm c/r^2$,
where we have assumed that the constant of integration $c$ is positive.
This relation can be written as
\be
\phi_r=\pm \frac{c}{\sqrt{r^4-\lx c^2}}.
\label{stat2} \ee
In ref. \cite{dvalex1} this solution was associated with the phenomenon of classicalization.
It was interpreted as a field configuration induced by a $\delta$-function source resulting from the large concentration of
energy within a small region of space around $r=0$.
For $\lx<0$ the solutions for both signs extend down to $r=0$.  For $\lx>0$ they display a
square-root singluarity at $r_s=\lx^{1/4} c^{1/2}$. They can be joined smoothly in order to create
a continuous double-valued function of $r$ that extends from infinite $r$ to $r_s$ and back out to infinity.

Exact dynamical solutions, of the form $\phi=\phi(z)$, with $z=r^2-(t-t_0)^2$, can also be found.
They are described by eqs. (\ref{ex1}), (\ref{ex2}), with
$h(z)$ satisfying
\be
z\, h'(z)+6\lx \, z\, h(z)^3+ 2 h(z)=0.
\label{hzeq} \ee
The solutions of this equation are
 \be
h(z)=\pm\frac{1}{\sqrt{c z^4-4\lx z}}.
\label{hz2}
\ee
Requiring that the solution remain real for $r\to\infty$ imposes $c>0$.
For both signs of $\lx$, the two solutions display square-root
singularities at the value $z_s=r^2_s-(t_s-t_0)^2$ that satisfies $z^3_s=4\lx/c$.

An interesting question is whether the analytical solutions we described capture some features of the evolution
of the incoming spherical wavepacket in the scattering problem. Unfortunately, we have not found any concrete evidence
for this possibility.
In ref. \cite{heisenberg}, solutions analogous to eqs.  (\ref{ex1}), (\ref{ex2}), ({\ref{hz2})
were employed in order to describe shock waves in the context of lower-dimensional field theories. The
shock waves we shall encounter in the following sections can be fitted by functions with square-root singularities, but with
coefficients that do not match the ones deduced from eq. (\ref{hz2}).
It seems that the connection of the known static or dynamical analytical solutions with the phenomenon of classicalization must
be considered rather sketchy. However, it is possible that richer analytical solutions may make the connection more concrete.

\section{Numerical method}\label{numer}

The numerical integration of the equations of motion is achieved through the implementation of a variant of
the leap-frog scheme. We have found that the most efficient form of the equations is given by eqs. (\ref{eomcons}), (\ref{eomconsd}).
The discretized version that we employ is
\be
\frac{\Dr}{\Dt} r_i^2\left[G\left(U_i^{j+1},V_i^{j+1}\right)-G\left(U_i^{j-1},V_i^{j-1}\right) \right]
=r^2_{i+1}F\left(U^j_{i+1},V^j_{i+1} \right)-r^2_{i-1}F\left(U^j_{i-1},V^j_{i-1} \right),
\label{eomdisc1} \ee
with $U=\phi_t$, $V=\phi_r$. The functions $F$ and $G$ are
$F(U,V)=U(1-\lx U^2 +\lx V^2)$, $G(U,V)=V(1-\lx U^2 +\lx V^2)$ for the first model and
$F(U,V)=U(1-\lx U^2 +\lx V^2)^{-1/2}$, $G(U,V)=V(1-\lx U^2 +\lx V^2)^{-1/2}$ for the second.
The index $j$ determines the discretized values of the time coordinate $t$, while the index $i$ the discretized values of the
radial coordinate $r$. The above equation must be complemented by an equation that enforces the condition
$\partial \phi_t/\partial r=\partial \phi_r/\partial t$. Its discretized version is
\be
U^{j}_{i+1}-U^{j}_{i-1} =\frac{\Dr}{\Dt}\left( V^{j+1}_{i}-V^{j-1}_{i} \right).
\label{eomdisc2} \ee
In practice, we solve eq. (\ref{eomdisc2}) for $V^{j+1}_i$, which we then substitute in eq. (\ref{eomdisc1}). The solution of the
resulting algebraic equation leads to the determination of $U^{j+1}_i$ in terms of known values at earlier times.
The boundary conditions at the ends of the radial interval are not of high importance, as our
analysis is restricted to time intervals during which the field is negligible there.

\begin{figure}
\includegraphics[width=140mm,height=80mm]{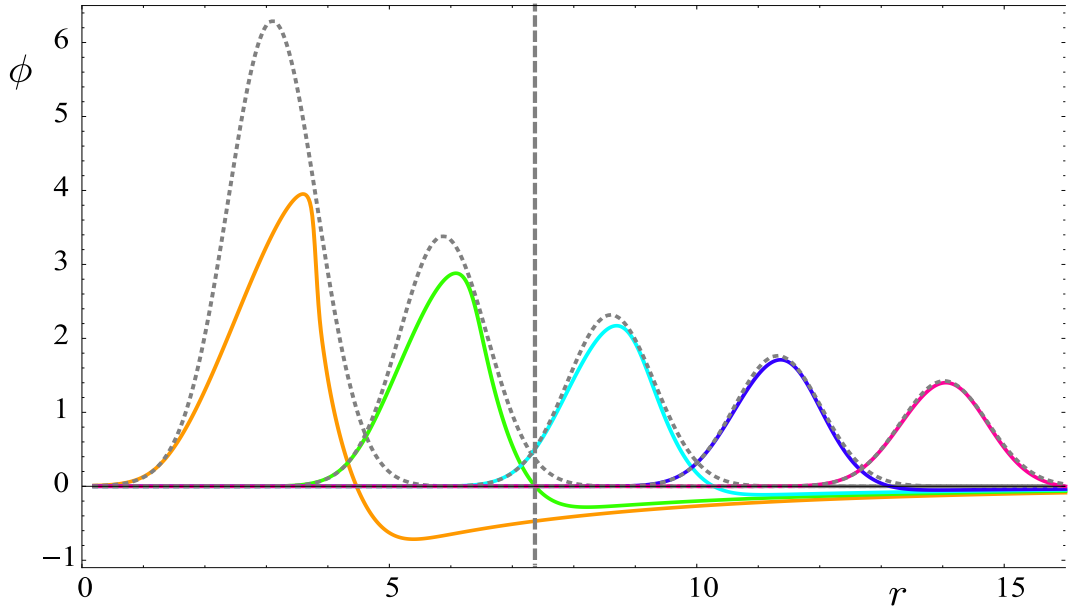}
\caption{The nonlinear wavepacket at various times (solid lines) vs. the linear wavepacket (dotted lines), in the context of the
DBI theory, described by eq. (\ref{lagrangian}),  with $\delta_2=1$, $L_*=1$. The initial wavepacket has $A=20$, $a=1$.
The vertical dashed line denotes the classicalization radius.
}
\label{fig1}
\end{figure}

The above scheme provides an accurate description of the evolution of the field configuration, starting from
an initial condition given by eq. (\ref{wave}) with $r_0\gg r_*\gg L_*$.
The initial stage is simple, as the equation of motion is essentially the wave equation. The evolution consists
mainly of the displacement of the peak of the wavepacket towards smaller values of $r$, with an increase of its amplitude.
The role of the higher-derivative terms becomes important during the later stages, when the wavepacket approaches the
classicalization radius $r_*$.
The most problematic
aspect for our analysis is that continuous real solutions may cease to exist at some stage of the evolution. There are three
reasons for this behavior:
\begin{itemize}
\item
At some stage the solution develops a
shock front. From this point on, the numerical integration cannot be continued, as the evolution of the shock
depends on additional physical assumptions about its nature. The equation of motion becomes singular at the location of
the shock and the resulting discontinuities in the field configuration, or its derivatives, cannot be determined without extra
input. We do not attempt to impose additional conditions, so as to continue the integration, for two reasons: Firstly, our
intuition on the physical properties of this system is limited; secondly, the scattering properties of the solution
can be established already before the appearance of the shock front.
We also mention at this point that the numerical solution is slightly better behaved near the
shock front, if one employs the quantities $F$ and $G$, appearing in eqs. (\ref{eomdisc1}), (\ref{eomdisc2}), as
independent variables, instead of $U$ and $V$. The reformulation of the system (\ref{eomdisc1}), (\ref{eomdisc2}) is
straightforward. Our results have been obtained and cross-checked through both formulations.
\item
At some time a real solution ceases to exist within a certain range of $r$. This possibility is already apparent in the
analytical solutions of the equations of motion that we discussed
at the ends of sections 2 and 3.  We encounter specific examples in the numerical analysis in the following section.
In the final section we provide an intuitive understanding of the origin of this behavior in the context of
toy models.
\item
The third type of complication occurs at the time when the partial differential equation switches type within a range of
$r$. This change is indicated by the discriminant $\Delta$ approaching zero from positive values. For $\Delta <0$ the
equation becomes elliptic and its solution requires (Dirichlet or Neumann)
boundary conditions on a closed contour around the region of interest.
Unfortunately, the scattering problem that we are considering cannot provide such conditions,
as it is set up through Cauchly boundary conditions at the initial time. For $\Delta <0$ the
coordinate $t$ becomes essentially spatial. Boundary conditions on a closed contour would require the values of
$\phi$ or its derivatives at times later than the time of interest. In the standard (and rather limited) studies of
partial differential equations of mixed type, the boundary conditions are imposed on appropriate boundaries so that
a unique solution exists. In the case we are considering this is not possible.
\end{itemize}

The above complications indicate that the idealized scattering problem that we are considering is not always well
posed mathematically. The most disturbing issue is the inability to find a solution in the range of $r$ where the equation becomes elliptic.
A reformulation of the problem is needed in order to overcome this difficulty.
Despite these misgivings, a solution of
the equation of motion over sufficiently long time intervals, so as to probe the region of classicalization,
is possible in many cases.
In the following section we present explicit examples.

\begin{figure}[t]
\includegraphics[width=140mm,height=80mm]{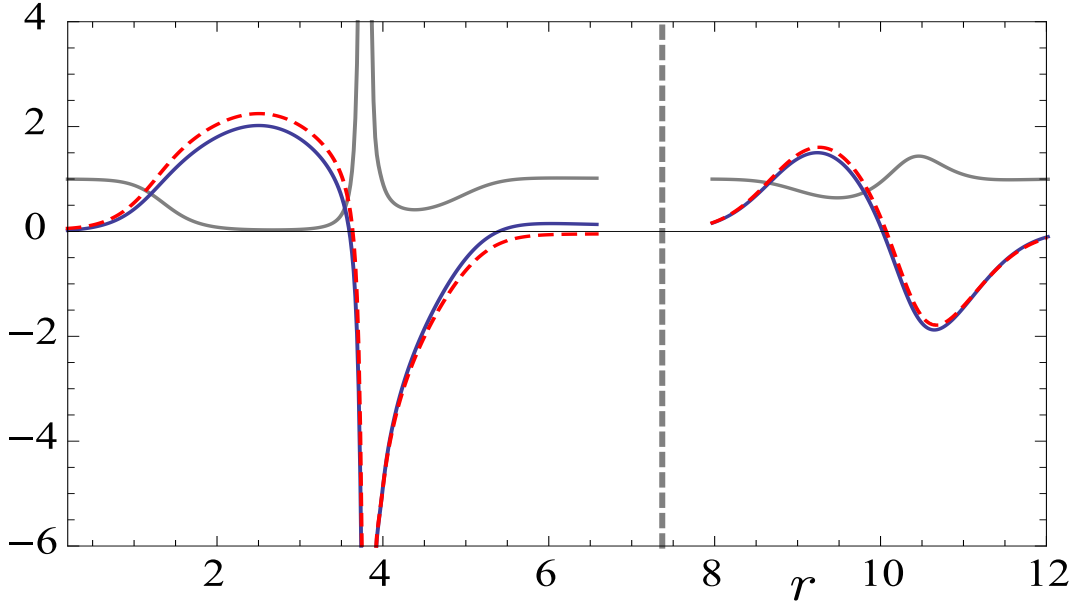}
\caption{The derivatives $\phi_t$ (dashed) and $\phi_r$ (solid) of the
nonlinear field, and the discriminant $\Delta$ of eq. (\ref{discr})
(solid grey), at two different times, before and after the crossing of the classicalization radius.
The model is the same as in fig. \ref{fig1}.
The vertical dashed line denotes the classicalization radius.
}
\label{fig2}
\end{figure}

\section{Numerical results} \label{resul}

We have found that the solutions with the most regular behavior are obtained in the context of the DBI model, described by the
Lagrangian (\ref{lagrangian}), with $\delta_2=1$ and, therefore, $\lx>0$.
In fig. \ref{fig1} we depict the solution of the equation of motion at various times. We have normalized all dimensonful quantities
with respect to the fundamental scale of the theory. This is equivalent to setting $L_*=1$. We have considered an initial
spherical wavepacket with a width equal to the fundamental scale ($a=1$), as the preparation of
a narrower configuration would presuppose that scales below $L_*$ can be probed. The amplitude of the configuration must
be much larger than 1 in order for the classicalization radius to satisfy $r_*\gg L_*$. We use $A=20$.
The solid lines denote the nonlinear field $\phi(t,r)$
resulting from the integration of eq. (\ref{eomcons}). The dotted lines denote the evolution of the initial configuration $\phi_0(t,r)$
if the nonlinear terms are neglected ($\lx=0$). In this case the equation of motion is the wave equation in spherical
coordinates, and eq. (\ref{wave}) provides its exact solution.

As expected, the nonlinear and linear field configurations coincide at early times, when the peak of the wavepacket
is located at distances much larger than $r_*$.
The nonlinear configuration is subject to visible deformations when the peak approaches $r_*$. The deformations
become significant when the wavepacket crosses inside the classicalization radius. The final configuration
in fig. \ref{fig1} depicts a wavepacket whose front end is located at distances below $L_*=1$. Its height is
reduced significantly in relation to the linear configuration. Its back end develops a long tail that extends far beyond
the classicalization radius. The last important feature of the configuration is that the part behind the peak is
very steep, essentially vertical. This indicates the apperance of a shock front, where the field derivatives become
singular.

In fig. \ref{fig2} we display some characteristics of the solution before and after the $r_*$-crossing. The dashed and
solid lines that almost coincide depict the derivatives $\phi_t$ and $\phi_t$, respectively. The grey solid line depicts
the discriminant $\Delta$ of eq. (\ref{eomsph}), given by eq. (\ref{discr}).
The form of the field at early times, seen on the right of the vertical dashed line marking the classicalization radius, is typical of
an incoming spherical wavepacket. The derivatives $\phi_t$, $\phi_r$ are almost equal, while the
discriminant deviates only slightly from 1.
The late-time configuration, located within the classicalization radius, has a significantly different form. Within
a small $r$-interval around the shock front, the
radial derivative $\phi_r$ becomes negative, with an absolute value that becomes very large. The time-derivative
$\phi_t$ has a similar form. At large $r$, both $\phi_t$, $\phi_r$ approach zero, having opposite signs. This is
an important feature that we shall discuss in detail below. The discriminant displays three types of behavior:
In the regions where $\phi_t$, $\phi_r$ are small, it takes values very close to 1, as the nonlinear terms in
the equation of motion are subleading. At the location of the shock front, it becomes very large. In the remaining
regions it approaches zero. This last feature is clearly visible in the region $2\lta r \lta 3.5$. It indicates that
the partial differential equation (\ref{eomsph}) becomes essentially parabolic within this region.
For wavepackets with larger amplitude $A$, the transitions in the values of $\Delta$ between 0, 1 and very large
values at the shock front become very sharp.

In fig. \ref{fig3} we display some additional features of the field configuration before and after the $r_*$-crossing.
The solid lines depict the field. The late-time configuration is characterized by a very long tail that extends
well beyond the classicalization radius. The dashed lines depict the product $4 \pi r^2 \rho$, with the
energy density $\rho$ given by eq. (\ref{density}). The total energy of the configuration,
corresponding to the area below the dashed lines, remains constant during the evolution. Actually, the
conservation of the energy is one of the criteria that we employ in order to estimate the accuracy of the
numerical integration. For both configurations depicted in fig. \ref{fig3}, the energy is distributed roughly equally between
the front and back end of the wavepacket. When the shock front develops, the energy density becomes strongly
peaked at its location. The long tail of the late-time configuration carries very little energy, as both
$\phi_t$, $\phi_r$ are very small there.

\begin{figure}[t]
\includegraphics[width=140mm,height=80mm]{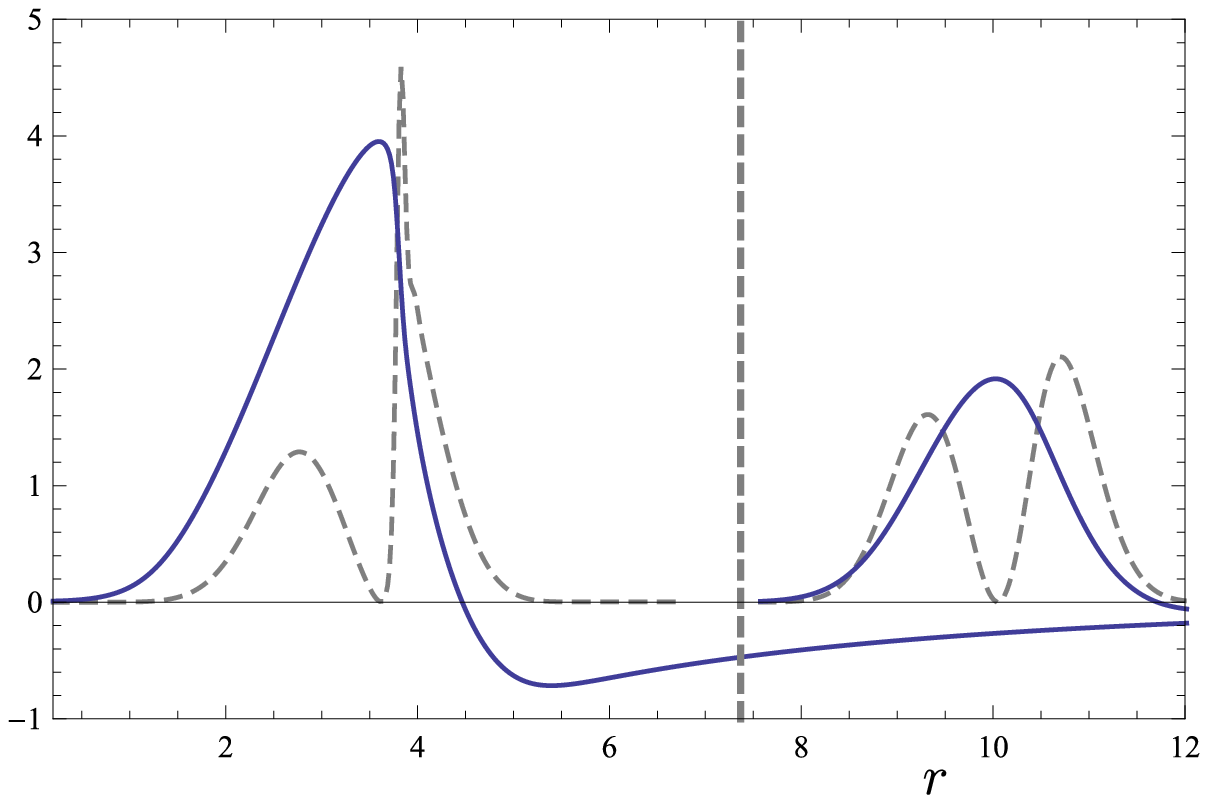}
\caption{
The nonlinear field $\phi$ (solid) and the product $4\pi r^2\rho$, with $\rho$ the energy density (dashed).
The model is the same as in fig. \ref{fig1}.
The vertical dashed line denotes the classicalization radius.
The energy density is multiplied by $5\times10^{-4}$. }
\label{fig3}
\end{figure}

The properties of the tail appearing at late times in figs. \ref{fig1}--\ref{fig3} are the main observable
consequence of the classicalization scenario. In the region within which the tail develops, the equation of motion
is essentially the wave equation, as $\phi_t$, $\phi_r$ are very small and the nonlinear corrections negligible.
For this reason we expect that the field can be expressed as
\be
\phi(t,r)=\frac{1}{r}\left[f(r+t)+g(r-t) \right].
\label{wavesol} \ee
The unknown functions $f$ and $g$ can be determined from our numerical solution through the relations
\begin{eqnarray}
f'(r+t)&=&\frac{1}{2} \left[r(\phi_r(t,r)+\phi_t(t,r))+\phi(t,r) \right]
\label{fpr}\\
g'(r-t)&=&\frac{1}{2} \left[r(\phi_r(t,r)-\phi_t(t,r))+\phi(t,r) \right],
\label{gpr}
\end{eqnarray}
where the primes denote derivatives with respect to the arguments.
Our results indicate that to a very good accuracy $f'(r+t)=0$, so that we can set consistently
$f(r+t)=0$ in eq. (\ref{wavesol}). Thus the tail of the field represents an outgoing
configuration, which can be interpreted as a result of the scattering of the initial incoming wavepacket.
We have also determined the form of the derivative of the function $g(r+t)$, by using data at different times.
The result is depicted in fig. \ref{fig4}, where we have indicated parts of the curve computed at different times.
They all form a continuous curve, with overlapping parts that are consistent with each to a high accuracy.
Within the tail, the field $\phi$ is a monotonically increasing function of $r+t$, starting from negative values
and vanishing for large arguments. We have performed various fits of the curve depicted in fig. \ref{fig4}.
The best one involves a combination of two terms  of the form $A/(r-t+c)^n$, with $A$, $c$, $n$
determined by the data. Even though the data do not provide a unique answer, we find that a combination of
powers with $n\sim 3-5$ gives the best fit.
This behavior is consistent with the analytical results of ref. \cite{dvpirts} for the theory
described by the Lagrangian density (\ref{lagrangian0}).

\begin{figure}[t]
\includegraphics[width=140mm,height=80mm]{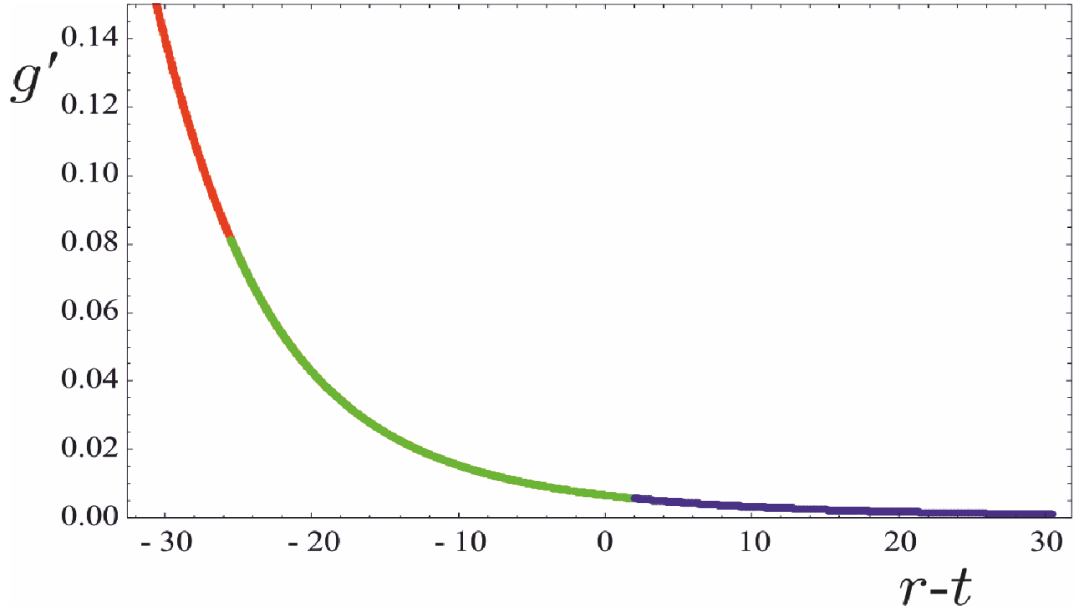}
\caption{The derivative of the function $g(r-t)$ appearing in the asymptotic form of the field (eq. (\ref{wavesol})).
The model is the same as in fig. \ref{fig1}. }
\label{fig4}
\end{figure}

The DBI model with $\delta_2=-1$ and $\lx<0$, with an initial condition given by eq. (\ref{wave}), displays the second of the
problems we described at the end of the previous section. A real solution
over the whole range of $r$ cannot be found after a certain time during the very early stages of the evolution.
The problem appears near the extrema of
$\phi_r$, where $1+\lx \phi_r^2$ vanishes. It is important to note that this term multiplies $\phi_{tt}$ if the equation
of motion is written in the form (\ref{eomsph}). In the following section we investigate the origin of this
behavior.

We next turn to the quartic model described by the Lagrangian (\ref{lagrangian0}). The most regular behavior
is obtained for the model with $\delta_1=-1$ and, therefore, $\lx<0$. The evolution of the initial wavepacket
is similar to that shown in fig. \ref{fig1}. In figs. \ref{fig5}, \ref{fig6} we depict
the solution at some early stage and at the latest time at which the solution is hyperbolic over the whole
range of $r$ ($\Delta >0$ everywhere).
The form of the solution is very similar to that depicted in figs. \ref{fig2}, \ref{fig3} for the DBI model.
In contrast to the
DBI model with $\lx>0$, the evolution at later times takes the system into the region where the equation of motion becomes
elliptic ($\Delta<0$) within a certain $r$-interval. As we explained in the previous section,
the boundary conditions provided by the
scattering problem are not appropriate for the determination of a unique solution of the equation of motion
when it becomes of mixed type (elliptic within a certain region, while staying hyperbolic in other regions).
We find that instabilities tend to develop during the
numerical integration when $\Delta$ becomes negative.
The accuracy of the integration remains high during an initial stage at which the equation of motion is of mixed type. It
diminishes gradually, until the numerical solution becomes unreliable at some time before the wavepacket reaches
the classicalization radius (denoted by the vertical dashed line in figs. \ref{fig5},\ref{fig6}).

Despite the above limitations, some useful information can be extracted from our solution.
It is obvious from fig. \ref{fig5} that the evolution of the wavepacket is similar to that in the DBI case with $\lx>0$.
The rear of the wavepacket becomes steeper and
a shock front starts developing.
This is indicated by $\phi_r$ becoming more negative. In the same time, a long tail is clearly visible at large values of $r$
in fig. \ref{fig6}.
The solid line in fig. \ref{fig5} represents the discriminant $\Delta$ given by
eq.  (\ref{discr0}). We have depicted the form of the wavepacket at the time when
$\Delta$ first reaches zero at a certain value of $r$.

Finally, the quartic model with $\lx >0$ displays the same behavior as the DBI model with $\lx<0$. A real solution
ceases to exist at some early stage in the evolution, at the time when the coefficient $1-3\lx\phi_t^2+\lx\phi_r^2$ of $\phi_{tt}$
in the equation of motion (\ref{eomsph0}) vanishes.

\begin{figure}[t]
\includegraphics[width=140mm,height=80mm]{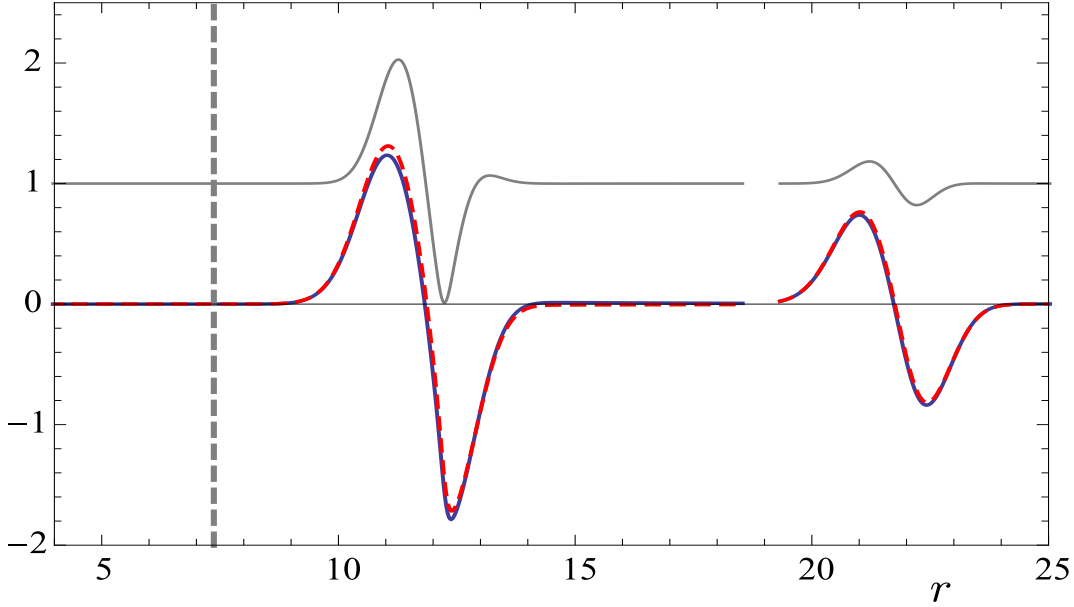}
\caption{The derivatives $\phi_t$ (dashed) and $\phi_r$ (solid) of the
nonlinear field, and the discriminant $\Delta$ of eq. (\ref{discr0})
(solid grey), at two different times, in the context of the
quartic theory, described by eq. (\ref{lagrangian0}),  with $\delta_1=-1$, $L_*=1$. The initial wavepacket has $A=20$, $a=1$.
The vertical dashed line denotes the classicalization radius.}
\label{fig5}
\end{figure}

\section{Discussion and conclusions}\label{concl}

Despite some inherent problematic issues, our analysis of the idealized scattering process with exact
spherical symmetry has provided concrete evidence for certain features of the classicalization scenario:
\begin{itemize}
\item
The classicalization radius of eq. (\ref{clr}) sets the scale for the onset of significant deformations of a
collapsing classical configuration with large energy concentration in a central region. This behavior is
consistent with the expectations of refs. \cite{dvalex1,dvpirts,dvali,dvalex2}.
\item
The equation of motion is a quasilinear partial differential equation of hyperbolic type at early times. At distances
comparable to the classicalization radius, the nonlinearities become significant and can change the equation type.
We discussed in detail an example in which the equation becomes
essentially parabolic in certain regions of space. This is consistent with the analysis of ref. \cite{brt}.
We also discussed an example in which the evolution makes the equation elliptic within a certain region. However,
a real solution cannot be found at late times within the elliptic region. It seems likely that
the initial conditions of the scattering process are not appropriate for the solution
of the mixed type equation.
\item
Shock fronts develop during the scattering process at distances comparable to
the classicalization radius, consistently with the expectations of ref. \cite{brt}.
\item
The most important observable feature of the classicalization process is the creation of an outgoing
field configuration that extends far beyond the classicalization radius. This feature
develops before the deformed wavepacket reaches distances of the order of the
fundamental scale $L_*$.
\end{itemize}

\begin{figure}[t]
\includegraphics[width=140mm,height=80mm]{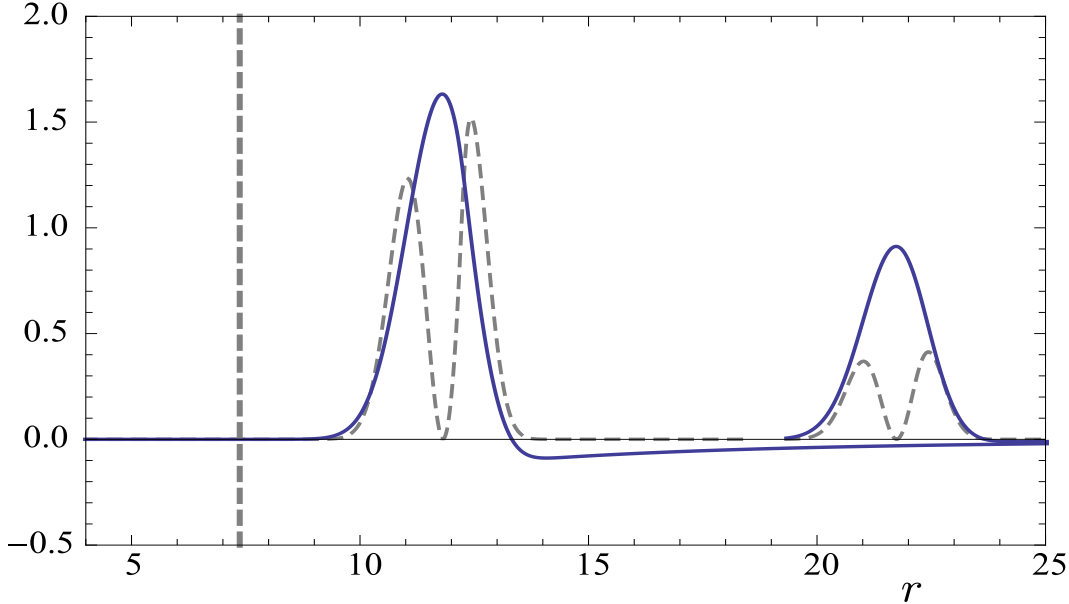}
\caption{
The nonlinear field $\phi$ (solid) and the product $4\pi r^2\rho$, with $\rho$ the energy density (dashed).
The model is the same as in fig. \ref{fig5}.
The vertical dashed line denotes the classicalization radius.
The energy density is multiplied by $5\times10^{-4}$.  }
\label{fig6}
\end{figure}

Our study has also revealed several novel issues related to classicalization:
\begin{itemize}
\item
Our numerical analysis in the context of the DBI model indicates that the collapsing wavepacket
can approach distances of the order of the
fundamental scale $L_*$ before strong scattering appears. However, this could be a special feature of the
DBI model. As we have seen, the equation of motion cannot become elliptic in this system, so that free
propagation of wavelike configurations is not eliminated.
\item
The scattering resulting from the early stages of the classicalization process seems to be minimal within our analysis. The tail of the
field configuration carries a negligible amount of energy because the corresponding modes are extremely soft.
Within the DBI model,
the bulk of the energy stored in the initial wavepacket can end up within a region of space with length scale comparable to
$L_*$. Similarly to the previous point, this behavior may be a special feature of the DBI model.
\item
Our analysis does not  provide evidence for the creation and subsequent decay of a quasistatic configuration that could be
identified with the classicalon of refs. \cite{dvalex1,dvpirts,dvali,dvalex2}. Instead, the classicalization scenario seems to
be a fully dynamical process, whose defining feature is that it is
characterized by a scale different from the naive length scale deduced from the higher-derivative
terms in the Lagrangian density. A curious fact is that the static classicalons seems to exist in the two cases
(quartic model with $\lx>0$ and DBI model with $\lx<0$) in which a dynamical solution ceases to exist at an
early stage of the evolution.
\item
The most unexpected issue that emerged in our analysis is that the scattering problem may not have
real solutions over the whole space within the context of the higher-derivative theories that support
classicalization.

\end{itemize}

In view of the above conclusions, a few more comments are in order concerning the accuracy of our analysis and the nature
of the solutions.
We have found that, in several of the versions of the equation of motion that we studied, an initial configuration representing a
collapsing spherical wavepacket cannot evolve beyond a certain time. A real solution of the equation of motion ceases to
exist for the initial condition of eqs. (\ref{wave}),(\ref{init}). It is always possible that the inability to find a solution
may be caused by the scheme we employed for the numerical solution of the equation of motion.
For this reason, apart from the explicit leap-frog method we described in section 4, we reproduced our results
through a second integration method, based on an implicit Crank-Nicolson scheme.
The agreement is very good in the regions in which a solution exists, while both methods indicate the absence of
a real solution in the same ranges of $t$ and $r$.

We would like to have a better understanding of the two cases in which
a real solution
ceases to exist at some early stage in the evolution, at the time when the coefficient of $\phi_{tt}$
in the equation of motion (\ref{eomsph0}) vanishes. These are
the quartic model with $\lx >0$ and the DBI model with $\lx<0$.
It must be emphasized that this behavior is not associated with the presence of square-root singularities,
similar to the ones appearing in the analytical solutions discussed at the ends of sections 2 and 3.
The values of $\phi_t$, $\phi_r$ are finite at the time at which a real solution ceases to exist.
We can develop intuition on the problem by considering
simple toy models that display similar behavior and can be solved explicitly. The simplest one is a harmonic oscillator of unit mass and
non-standard kinetic terms.

The analogue of the quartic model we discussed in section 2 is described by
a Lagrangian
\be
L=\frac{1}{2} \dot{x}^2- \frac{1}{4}\delta \,\xi\, \dot{x}^4-\frac{1}{2}\omega^2 x^2,
\label{toy1} \ee
with $\delta=\pm1$ and $\xi>0$.
The equation of motion is
\be
(1-3\,\delta\,\xi\,\dot{x}^2)\ddot{x}+\omega^2 x=0.
\label{eomtoy1} \ee
Its first integral is the conserved energy
\be
E=\frac{1}{2}\dot{x}^2-\frac{3}{4}\delta\, \xi\, \dot{x}^4+\frac{1}{2}\omega^2 x^2 =\frac{1}{2}\omega^2 x_0^2,
\label{energytoy1} \ee
where we have assumed that initially the particle is at rest at $x=x_0$.
For this initial condition, the above equation gives
\be
\dot{x}^2=\frac{1-\sqrt{1-6\,\delta\, \xi\, \omega^2(x_0^2-x^2)}}{3\,\delta\,\xi}.
\label{soltoy1} \ee
For $\delta=-1$ a solution exists at all times, describing a particle oscillating around the minimum of the potential at $x=0$.
For $\delta=1$ a real oscillating solution exists at all times if
$x^2_0<(6\,\xi\,\omega^2)^{-1}$. However, for $x^2_0>(6\,\xi\,\omega^2)^{-1}$ a real solution
does not exist below the point with
$x^2=x^2_0-(6\,\xi\,\omega^2)^{-1}$. This is reached at the time when the coefficient of $\ddot{x}$ in eq. (\ref{eomtoy1})
vanishes. We observed very similar behavior in the solutions described in section 5.

The crucial property of the Lagrangian (\ref{toy1}) with $\delta=1$ is that the corresponding energy, given by eq. (\ref{energytoy1}), is
bounded from above when considered as a function of $\dot{x}$.
\footnote{The fact that the energy (\ref{energytoy1}) is unbounded from below for $\dot{x}\to \infty$ is not relevant for the classical
evolution that we consider: The system starts with $\dot{x}=0$ and the classical evolution breaks down at the
finite value of $\dot{x}$ that
corresponds to the "top of the hill" of the kinetic energy.}
If the evolution starts at a suffciently large value of $x$, the
initial potential energy can be larger than the maximal possible kinetic energy. As a result, the conservation of energy cannot
be satisfied at some point during the evolution of the particle towards the minimum of the potential, and a real solution
ceases to exist. The maximal value of the kinetic energy is obtained for $\partial E/\partial \dot{x}=\dot{x}(1-3\delta \xi \dot{x}^2)=0$.
This explains why a real solution does not exist beyond the time at which the coefficient of $\ddot{x}$ in eq. (\ref{eomtoy1})
vanishes. It is also interesting to consider the momentum conjugate to $x$, given by $p=\dot{x}(1-\delta \xi \dot{x}^2)$.
This expression has a maximum for $\dot{x}^2=1/(3\delta\xi)$, given by $\sqrt{\delta\xi}p_m=2/(3\sqrt{3})$.
There is no continuous real function $\dot{x}=\dot{x}(p)$, defined over the whole real axis for $p$, that behaves
like $\dot{x}\simeq p$ for $p\to 0$.

The field theoretical model of eq. (\ref{lagrangian0}) with $\delta_1=1$ and $\lx>0$ has very similar features with the toy model.
The energy density of eq. (\ref{density0}) has a maximum as a function of $\phi_t$ at the point at which the
coefficient of $\phi_{tt}$ in eq. (\ref{eomsph0}) vanishes. This occurs at the time at which $1-3 \lx \phi_t^2+\lx\phi_r^2$
becomes zero for a certain value of $r$ and
the numerical integration stops converging. The origin of the problem is now clear: The energy cannot be conserved
beyond this point and a real solution does not exist any more.
The momentum density conjugate to $\phi$, given by $\pi=\phi_t(1-\phi_t^2+\phi_r^2)$, also has a maximum as
a function of $\phi_t$ when $1-3 \lx \phi_t^2+\lx\phi_r^2=0$, similarly to the toy model.

The analogue of the DBI model of section 3 is described by
a Lagrangian
\be
L=-\frac{1}{\delta\, \xi} \sqrt{1-\delta\,\xi\,\dot{x}^2}-\frac{1}{2}\omega^2 x^2,
\label{toy2} \ee
with $\delta=\pm1$ and $\xi>0$.
The equation of motion is
\be
\ddot{x}+(1-\delta\,\xi\, \dot{x}^2)^{3/2}\,\omega^2 x=0,
\label{eomtoy2} \ee
and the conserved energy
\be
E=\frac{1}{\delta\,\xi  \sqrt{1-\delta\,\xi\,\dot{x}^2}  }+\frac{1}{2}\omega^2 x^2 =\frac{1}{\delta\,\xi  }+\frac{1}{2}\omega^2 x_0^2,
\label{energytoy2} \ee
where we have assumed that initially the particle is at rest at $x=x_0$.
The above equation can be rewritten as
\be
\frac{1}{ \sqrt{1-\delta\,\xi\,\dot{x}^2}  }-1=\frac{1}{2}\delta\,\xi\,\omega^2 (x_0^2-x^2).
\label{soltoy2} \ee
For $\delta=1$ a solution exists at all times, describing a relativistic particle oscillating around the minimum of the potential at $x=0$.
 We observed very similar behavior in the solution depicted in figs. \ref{fig1}-\ref{fig3}, which exists at all times up to the formation
of the shock front.

For $\delta=-1$ a real oscillating solution always exists if
$x^2_0<(2\,\xi\,\omega^2)^{-1}$. However, for $x^2_0>(2\,\xi\,\omega^2)^{-1}$ a real solution
ceases to exist when the particle reaches the point with
$x^2=x^2_0-(2\,\xi\,\omega^2)^{-1}$, at which $\dot{x}$ diverges. The origin of the problem can again be traced
to the fact that the kinetic energy in eq. (\ref{energytoy2}) is bounded from above when considered as a function of
$\dot{x}$. The maximum occurs for $\dot{x}\to \infty$. The conjugate momentum $p=\dot{x}/\sqrt{1+\xi \dot{x}^2}$
also has a maximal value, equal to $1/\sqrt{\xi}$, obtained for $\dot{x}\to \infty$. The corresponding field theory is the DBI model of eq. (\ref{lagrangian}) with
$\delta_2=-1$ and $\lx<0$. The partial derivative with respect to $\phi_t$  of the energy density of eq. (\ref{density}) vanishes
at the point
where $1+\lx \phi_r^2$, the coefficient of $\phi_{tt}$ in eq. (\ref{eomsph}), becomes zero. The partial derivative with
respect to $\phi_t$ of the momentum
density $\pi=\phi_t/\sqrt{1-\lx\phi_t^2+\lx\phi_r^2}$ also vanishes at the same point.
These properties are responsible for the absence of a real solution of eq. (\ref{eomsph}) beyond this point,
as the classical evolution cannot conserve the total energy of the system.

In summary, our analysis has confirmed the existence of nontrivial physical behavior at distances much larger than the
fundamental length scale in theories that support classicalization. It has also revealed some peculiar features of
certain underlying theories,
which may lead to the absence of real solutions of the classical equations of motion.
It remains to be seen whether the breakdown of the classical evolution is a pathology of such
theories, or whether it signals the transition to a regime in which the role of quantum physics is dominant.
The one-dimensional toy models we considered in this section may provide crucial intuition on this issue.

\section*{Acknowledgments}
We would like to thank N. Brouzakis for collaboration during the early stages of this work and many useful discussions.
We would also like to thank G. Dvali for useful discussions.
This research has been supported in part by
the ITN network ``UNILHC'' (PITN-GA-2009-237920).
This research has been co-financed by the European Union (European Social Fund – ESF) and Greek national
funds through the Operational Program ``Education and Lifelong Learning" of the National Strategic Reference
Framework (NSRF) - Research Funding Program: ``THALIS. Investing in the society of knowledge through the
European Social Fund".

\end{document}